\documentclass[%
reprint,superscriptaddress,amsmath,amssymb,aps,prl,longbibliography
]{revtex4-2}

\usepackage{graphicx}
\usepackage{dcolumn}
\usepackage{bm}
\usepackage{amssymb}
\usepackage{amsmath}
\usepackage{float}
\usepackage{subfigure}
\usepackage{xcolor}
\usepackage{color, soul}
\usepackage{tabularx}
\usepackage{array}

\usepackage[flushleft]{threeparttable}
\usepackage{amsthm}
\usepackage{amsfonts}
\usepackage{braket}
\usepackage{natbib, hyperref}
\hypersetup{colorlinks=true,linkcolor=blue, filecolor=blue,urlcolor=blue,citecolor=blue,}

\usepackage [english]{babel}
\usepackage [autostyle, english = american]{csquotes}
\MakeOuterQuote{"}

\usepackage[version=4]{mhchem}
\usepackage{chemmacros}
\chemsetup{
    modules = all,
    formula = mhchem
}

\usepackage{chemformula}

\usepackage{soul}

\begin{document}

\pagestyle{plain}

\title{The case of the missing gallium vacancy in gallium arsenide: \\ A multiscale explanation}

\author{Leopoldo Diaz}\noaffiliation
\author{Harold P. Hjalmarson}\noaffiliation%
\author{Jesse J. Lutz}\noaffiliation
\author{Peter A. Schultz}
 \email{paschul@sandia.gov}
\affiliation{%
 Sandia National Laboratories, Albuquerque, New Mexico 87185, USA
}%

\begin{abstract}
Irradiation of gallium arsenide (GaAs) produces immobile vacancies and mobile interstitials. However, after decades of experimental investigation, the immobile Ga vacancy eludes observation, raising the question: Where is the Ga vacancy? Static first-principles calculations predict a Ga vacancy should be readily observed. We find that short-time dynamical evolution of primary defects is key to explaining this conundrum. Introducing a multiscale Atomistically Informed Device Engineering (AIDE) method, we discover that during the initial displacement damage, the Fermi level shifts to mid-gap producing oppositely charged vacancies and interstitials. Driven by Coulomb attraction, fast As interstitials preferentially annihilate Ga vacancies, causing their population to plummet below detectable limits before being experimentally observed. This innovative model solves the mystery of the missing Ga vacancy and reveals the importance of a multiscale approach to explore the dynamical chemical behavior in experimentally inaccessible short-time regimes.

\end{abstract}

\maketitle
Atomic defects play a significant role in semiconductor properties. Understanding their chemical behavior has been a central theme in several investigations \cite{zhang2015,mitterreiter2021,ogawa2022,na2024}. Defect physics in gallium arsenide (GaAs) has been intensively studied for many decades \cite{pons1980,pons_bourgoin1985,bourgoin1988,schultz2009}; however, despite the progress several open questions persist \cite{fleetwood2021}. Irradiation of GaAs by energetic particles---through deliberate ion implantation (to controllably dope a material), exposure to high-energy electrons, or ions in radiation environments (e.g., satellite electronics)---displaces atoms from their lattice positions leaving behind immobile vacancies and mobile interstitials. Identifying and characterizing these atomic defects and their subsequent chemical evolution (annealing), through defect reactions, is essential to developing the comprehensive understanding necessary to enhance and improve the longevity of semiconductor devices. Currently, our understanding of these processes in GaAs remains incomplete and insufficient. 

\begin{figure} [t!]
    \centering
    \includegraphics[width=\columnwidth]{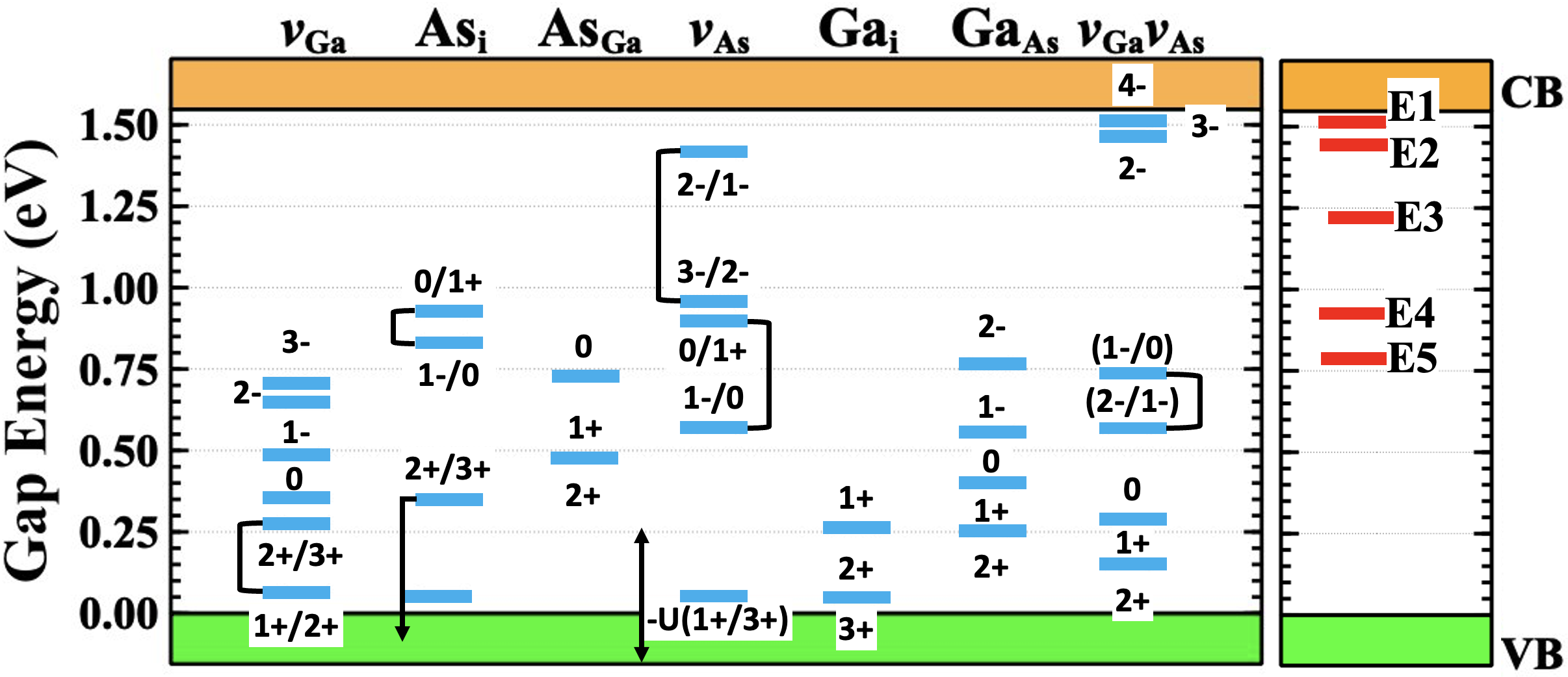}
    \caption{DFT-computed defect level diagram extracted from \cite{schultz2009}. Solid lines connect levels that exhibit -U behavior. Conduction (CB) and valence (VB) bands are orange and green. }
    \label{fig:fig1}
\end{figure}

A provisional assignment of GaAs defects has been established through a series of experimental and theoretical studies. Theory was pivotal in definitively identifying the technologically crucial EL2 in as-grown GaAs as the arsenic antisite $As_{Ga}$ (As atom occupying a Ga site) \cite{dabrowski1988,chadi1988}. Deep-level transient spectroscopy (DLTS) experiments on irradiated n-GaAs revealed a collection of defect levels, denoted E1-E2, E3, E4, and E5 \cite{pons_bourgoin1985}, at or above mid-gap. With limited theoretical support and capability, defect identification proved to be challenging for several decades. However, using density functional theory (DFT) modeling, Schultz and von Lilienfeld provided predictions for intrinsic point defects and defect levels in GaAs (Fig. \ref{fig:fig1}) \cite{schultz2009}. Theory, now equipped with this Rosetta Stone of GaAs defects, was instrumental in identifying several defects including the primary E1-E2, two distinct transitions of the same defect near the conduction band edge (CBE) \cite{pons_bourgoin1985,Loualiche1984,Corbel1988,lai1994}, as the divacancy ($v_{Ga}v_{As}$) \cite{schultz2015} and the E3 center as the As vacancy ($v_{As}$) \cite{schultz2009,schultz2015}. The E4 and E5 centers, located near mid-gap among a zoo of defects, remain unidentified and are likely due to a complex of point defects.

\begin{figure*}[t!]
    \centering
    \includegraphics[width=\textwidth]{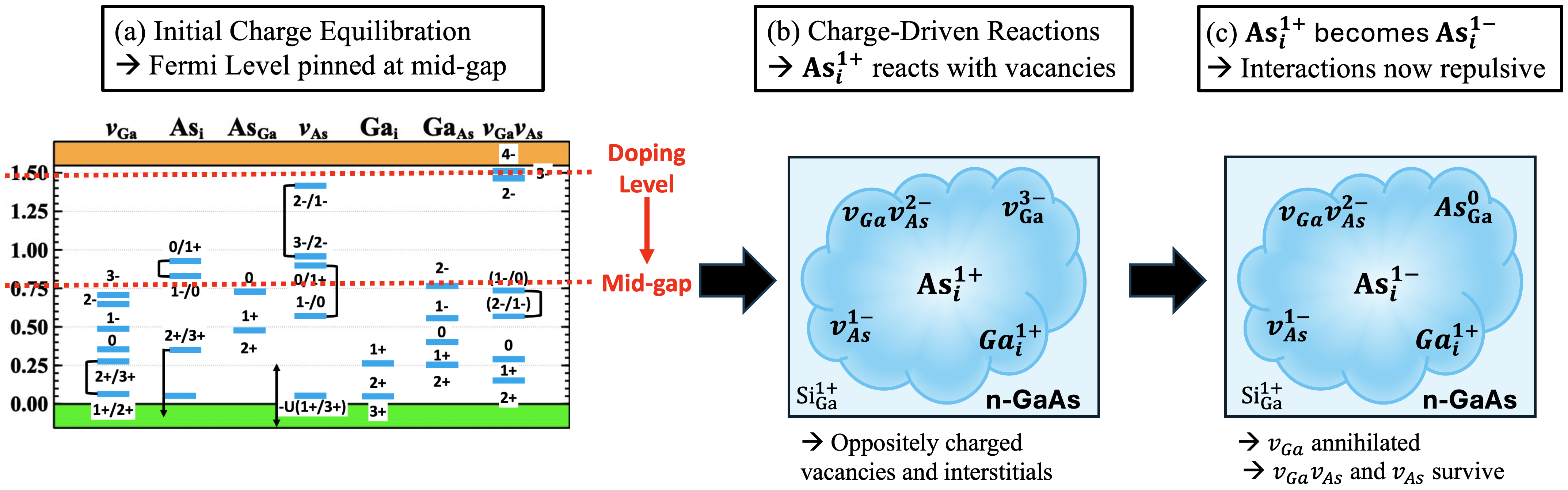}
    \caption{An illustrative description of the Conceptual Model used in this study showing (a) the initial charge equilibration (i.e., Fermi level stabilization at mid-gap), (b) the most populated charge state for each defect causing charge driven reactions, and (c) the dominant charge state of the surviving defects after the Ga vacancy is annihilated.}
    \label{fig:fig2}
\end{figure*}

Recently, a Laplace DLTS study \cite{taghizadeh2018} resolved the E3 peak into three distinct components---E3a, E3b, and E3c. The E3a (principal component) and E3c were identified as $v_{As}$ \cite{schultz2009} and the vacancy-Si pair \cite{schultz2022}, respectively. The E3b (CBE$-0.38$ eV) has not been theoretically identified but was determined to be intrinsic \cite{taghizadeh2018}. None of the defect identifications included the Ga vacancy ($v_{Ga}$). Despite theory predicting $v_{Ga}$, like $v_{As}$ and $v_{Ga}v_{As}$, to be stable \cite{baraff1985,mellouhi2005,northrup1993,schick2002}, immobile \cite{El-Mellouhi2006}, and have multiple charge states \cite{schultz2009,baraff1985,mellouhi2005,northrup1993,schick2002}, no experimental measurement has unambiguously observed the $v_{Ga}$. 

The absence of $v_{Ga}$ is a crucial missing link in understanding radiation damage in GaAs. It remains unclear whether this gap is due to a failure in understanding the nature of the initial displacement damage (as Frenkel pairs), some undiagnosed blindness of experimental probes to $v_{Ga}$, or a failure in the theoretical and experimental interpretation of $v_{Ga}$ stability and mobility. This understanding has eluded investigators for decades.

Our primary motivation is to confront this fundamental puzzle: Where is the missing $v_{Ga}$?  

Theory has reached a consensus that $v_{Ga}$ exists in a $3-$ charge state above mid-gap \cite{baraff1985,schultz2009}. Meanwhile, neither DLTS nor any other experimental technique has witnessed any firm indication of $v_{Ga}$, inferring that any transition levels must lie near or below mid-gap. This provides the first essential clue needed to explain the observational absence of $v_{Ga}$: it exists as a $3-$ defect above mid-gap.

The Conceptual Model of irradiated n-type (Si-doped) GaAs begins with standard displacement damage: energetic particles displace mobile interstitial
atoms ($Ga_i$ and $As_i$) from their lattice sites leaving behind immobile vacancies ($v_{Ga}$ and $v_{As}$) and divacancies ($v_{Ga}v_{As}$). Immediately following the primary displacement damage, electrons flow between defects. Mediated by charge carriers, electrons fill defect states deepest in the band gap until establishing an equilibrium Fermi level---charge equilibration (CE).

The Ga vacancy assumes a $3-$ charge state at mid-gap (Fig. \ref{fig:fig1}). The irradiation event generates sufficiently many Ga vacancies, each capturing three electrons, which capture sufficient electrons from higher defect levels in the band gap to shift the Fermi level from the Si-doping level to mid-gap. This Fermi level shift to mid-gap alters all defect charge states, producing negatively-charged vacancies and positively-charged interstitials [Fig. \ref{fig:fig2}(a)].

\begin{figure}[t!]
    \centering
    \includegraphics[width=\columnwidth]{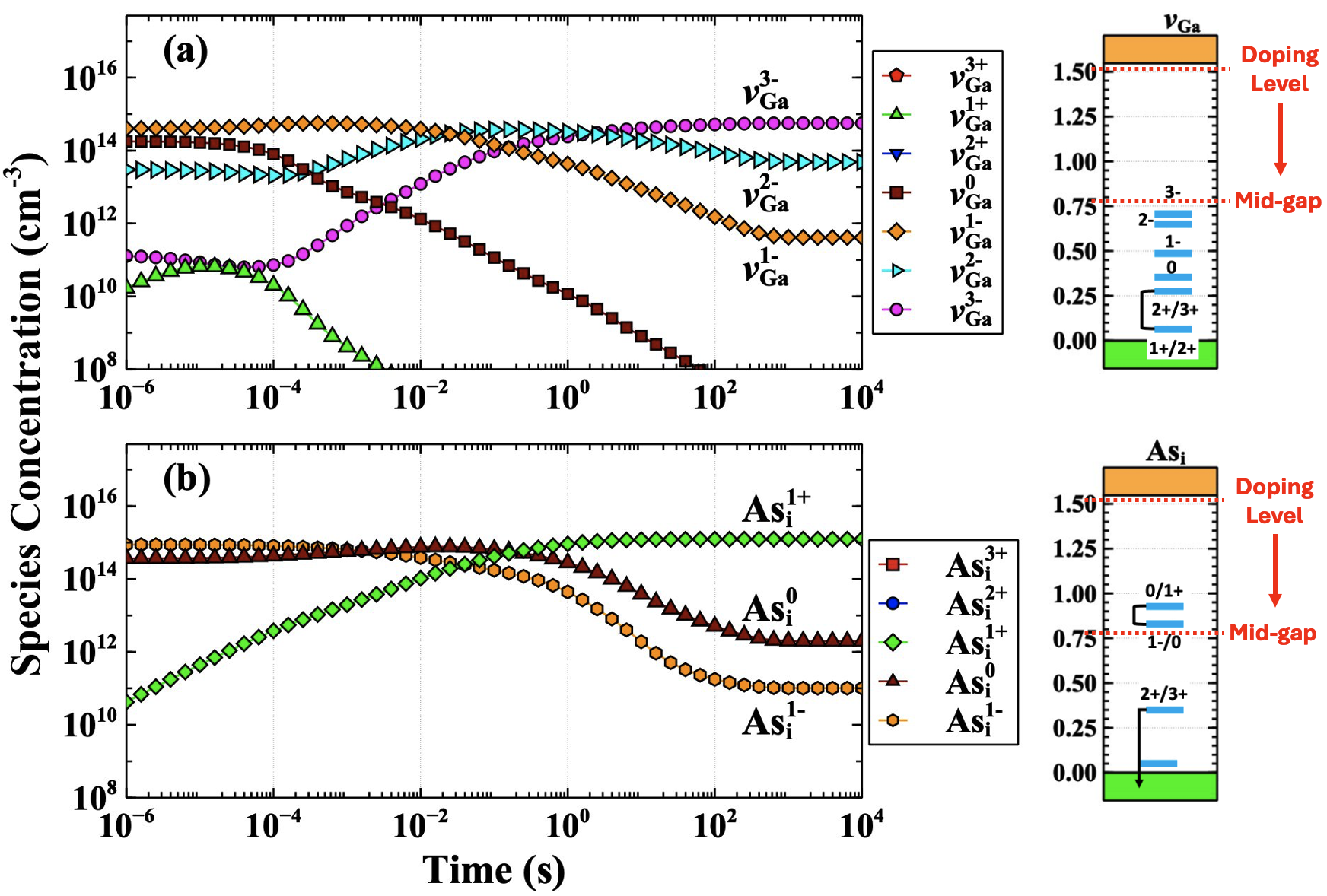}
    \caption{Calculated species concentration for (a) $v_{Ga}$ and (b) $As_{i}$ charge states during charge equilibration. Charge states not shown exist below $10^8$ cm$^{-3}$. (c) Defect levels of $v_{Ga}$ and $As_i$ with Fermi level shift to mid-gap ($E_F\approx0.76$ eV).}
    \label{fig:fig3}
\end{figure}
Theory predicts that $As_i$ becomes positively charged below mid-gap ($E_F=0.76$ eV). This prediction is significant because both experimental \cite{bourgoin1988} and theoretical \cite{schultz2009} studies have asserted that $As_i$ undergoes fast, athermal diffusion \cite{bourgoin-corbett} as a positive ion: a fast $As_i$ dominates the first stage of defect reactions. The $As_i^{1+}$ are attracted by all negatively charged vacancies, but the strongest Coulomb attraction is from $v_{Ga}^{3-}$. The $As_i^{1+}$ preferentially annihilates $v_{Ga}^{3-}$ to form $As_{Ga}^0$ ($As_{i}^{1+}+v_{Ga}^{3-}\rightleftarrows As_{Ga}^0 + 2e^-$) causing its population to plummet. With $v_{Ga}^{3-}$ eliminated, the Fermi level becomes unpinned, rising so that $As_i^{1-}$ becomes its dominant charge state. The $As_i^{1-}$ interaction with all negatively charged vacancies becomes repulsive, shutting down any further annihilation by $As_i$. This annihilation, consistent with experimental observations, suggests that only As-site vacancies ($v_{As}$ and $v_{Ga}v_{As}$) survive [Fig. \ref{fig:fig2}(c)].

The $v_{Ga}$ are annihilated too quickly to be seen by experimental measurements; the fading $v_{Ga}$ are invisible to experiment. While this Conceptual Model presents a plausible chain of events inferred from the DFT data, it invokes kinetic processes whose effects cannot be assessed with stationary-static DFT calculations. To account for the kinetic processes, we develop an Atomistically Informed Device Engineering (AIDE) method, which uses defect properties from DFT and experimental observations/measurements. 

The multiscale AIDE method is very powerful and encompasses the necessary dynamical physical phenomena to explain why $v_{Ga}$ is invisible to experimental probing.

\textit{Device Model.} Using the Radiation Effects in Oxides and Semiconductors (REOS) software package \cite{hjalmarson2003}, the dynamical nature of defects in irradiated GaAs is studied at 300 K. Unlike standard commercial device simulation codes, REOS is populated with DFT-calculated defect levels and available experimental data, creating a physically accurate treatment of dynamical systems. An important advantage is its ability to probe regimes---early-time behavior and extremely low populations---that are inaccessible to current experimental techniques.

A one-dimensional slab of length $2.1\times10^{-4}$ cm is used in the device simulations. The DFT-predicted effective band gap (1.54 eV, experimental value is 1.52 eV \cite{vurgaftman2001}) is used and populated with the computed defect levels (Fig. \ref{fig:fig1} \cite{schultz2009}). In addition to the simple intrinsic defects predicted by DFT, a Si-dopant is added to the GaAs sample at a concentration of $1.00\times10^{15}$ cm$^{-3}$, with two charge states: $Si_{Ga}^{1+}$ and $Si_{Ga}^0$, consistent with typical experimental doping concentrations ranging from $10^{15}-~10^{16}$ \cite{taghizadeh2018}. Chosen defect populations are $As_i=Ga_i=1.24 \times 10^{15}$, $v_{As} = v_{Ga} = 6.10 \times 10^{14}$, and $v_{Ga}v_{As} = 6.30 \times 10^{14}$ cm$^{-3}$ (criterion in End Matter). 

\textit{Charge equilibration (CE).} After primary displacement damage, electrons migrate throughout the system, filling the lowest states first, until achieving equilibrium---Fermi level stabilization---producing negatively-charged vacancies and positively-charged interstitials [Figs. \ref{fig:fig2}(a) and (b)]. In agreement with our Conceptual Model, CE results in the Fermi level shifting from the Si-doping level to mid-gap ($E_F\approx0.76$ eV). To confirm this shift, defect species populations were calculated for each defect and compared to one another; all dominant charge states must correlate with the same Fermi level. In Fig. \ref{fig:fig3}, species densities for all $v_{Ga}$ and $As_i$ charge states are shown. The model, consistent with DFT-predicted defect levels, identifies $As_i^{1+}$ and $v_{Ga}^{3-}$ as the dominant charge states. Not shown here, the model also correctly identifies $v_{As}^{1-}$, $Ga_i^{1+}$, and $v_{Ga}v_{As}^{2-}$ as the dominant defect charge states. Additionally, the calculated charge-state ordering for all defects agrees with the DFT-predicted levels, including those below measurable limits. The Fermi level collapses to mid-gap pinned by the $v_{Ga}$ ($3-/2-$) defect level.

Despite being able to predict each defect’s dominant charge state from the defect level diagram (Fig. \ref{fig:fig1}), performing CE dynamically with REOS serves as an additional metric to ensure the simulations physical accuracy. To enhance the physical realism of the CE process, specifically to achieve faster equilibration, an initial condition was implemented lowering the CE time from 1 to $>10^{-12}$ s (Supplemental Information Fig. S1). This initial condition produced identical results and is used for the remainder of this study. 

\textit{Coulomb-driven defect-defect reactions.} Fermi level stabilization established three negatively-charged vacancies ($v_{As}^{1-}$, $v_{As}v_{Ga}^{2-}$, and $v_{Ga}^{3-}$) and two positively-charged interstitials ($As_{i}^{1+}$ and $Ga_{i}^{1+}$) as the most populated defect species. Coulomb attraction between oppositely charged species triggers a sequence of defect-defect reactions that leads to the formation of: (i) new defects, (ii) point defect complexes, and/or (iii) healed lattice points via recombination. Like-charged species, conversely, repel one another. Since defect-defect reactions involve several possible outcomes, it is expected that charge state population sizes will change throughout this process and, in response to charge flow, cause the Fermi level to shift once again. Therefore, during this stage of our Conceptual Model [Fig. \ref{fig:fig2}(b)], two reaction types ensue: charge carrier capture/emission reactions and Coulomb-driven defect-defect reactions (diffusion-driven reactions).

Diffusion-driven reactions are governed by the diffusion coefficient in which the migration barrier energy ($E_m$) is essential for species diffusion in the simulation. In GaAs, vacancies are immobile \cite{mellouhi2006,mellouhi2007,schultz2009}; consequently, any chemical evolution must be mediated by the mobile As and Ga interstitials. However, $As_i^{1+}$ has a lower thermal migration barrier ($E_m^{As_i^{1+}}=0.50$~eV \cite{bourgoin1988} vs. $E_m^{Ga_i^{1+}}\approx1.0$~eV \cite{Malouin2007,schultz2009,Schick2011}) and diffuses athermally via the Bourgoin-Corbett mechanism \cite{bourgoin-corbett}. Any initial defect chemistry in GaAs will be dominated by the very fast $As_i^{1+}$ athermal diffusion.

Athermal processes (Bourgoin-Corbett), while understood conceptually, remains a challenge to describe \cite{koshka2004} and integrate into simulation. Athermal processes were not simulated directly in REOS. While some authors have treated it as an additive term to the thermal diffusion coefficient \cite{wampler2015}, we adopt an alternate approach and mimic the fast athermal diffusion effects by reducing $As_i^{1+}$ thermal migration barrier (0.50 eV \cite{bourgoin1988}) by 20\% to 0.40 eV.

Each reaction must produce a stable DFT-predicted product defect (Fig. \ref{fig:fig1}), thereby ensuring its physical presence in the material. Limiting the mobile species to $As_i^{1+}$ and adhering to the stability prerequisite limits the number of possible reactions and products formed. From weakest to strongest Coulomb attraction, the three reactions and their products are

\begin{reactions*}
    \hspace{1cm}&\llap{Reaction 1} & As_{i}^{1+} + $v$_{As}^{1-} & <=> z_{As}^{0} \\
    &\llap{Reaction 2} & As_{i}^{1+} +$v$_{Ga}$v$_{As}^{2-} & <=> z_{As}^{0} + $v$_{Ga}^{1-} \\
    &\llap{Reaction 3} & As_{i}^{1+} + $v$_{Ga}^{3-} & <=> As_{Ga}^{0} + 2e^{-}
\end{reactions*}
where Reaction 1 results in a healed As-lattice point ($z_{As}$), Reaction 2 produces $z_{As}^0$ and $v_{Ga}^{1-}$, and Reaction 3 forms $As_{Ga}^0$ with 2$e^-$ being released into the system. Reaction 3 is expected to form $As_i^{1+}+v_{Ga}^{3-}\rightleftarrows As_{Ga}^{2-}$, however, according to our defect stability prerequisite, $As_{Ga}^{2-}$ is predicted by DFT to be absolutely unstable \cite{schultz2009} and cannot be a forming product. 

Currently, the physical mechanism of Reaction 3---emission of multiple electrons in a single step---is not well understood. In REOS, multi-electron emission is represented as a series of intermediate charge-conserving reactions that involve the emission of a single electron:
\begin{reactions*}
    \hspace{0.75cm}&\llap{Reaction 4} & As_{i}^{1+} + $v$_{Ga}^{3-} &<=> As_{i}$-v$_{Ga}^{2-} \\
    &\llap{Reaction 5} & As_{i}$-v$_{Ga}^{2-} + h^+ &<=> As_{i}$-v$_{Ga}^{1-} \\
    &\llap{Reaction 6} & As_{i}$-v$_{Ga}^{1-} + h^+ &<=> As_{i}$-v$_{Ga}^{0} \\
    &\llap{Reaction 7} & As_i$-v$_{Ga}^{0} &<=> As_{Ga}^{0}
\end{reactions*}
where $As_{i}-v_{Ga}^{2-}$, $As_{i}-v_{Ga}^{1-}$, and $As_{i}-v_{Ga}^{0}$ are short-lived intermediates formed during the process. Reactions 4-6 occur rapidly, serve only to aid our simulation, and produce thermodynamically unstable product defects. These intermediate steps assure (i) the formation of a stable $As_{Ga}^0$ and (ii) the release of 2$e^-$ into the system. Physically, Reaction 7 has a separate mechanism in which $As_i$ must overcome an energy barrier before hopping into vacancy sites. Formation of a complex and overcoming of an energy barrier is true for all $As_i^{1+}$-vacancy reactions (End Matter and Supplementary Information).

Simulated reactions include Reactions 1-2 and 4-7. Notably, the Coulomb attraction for each reaction is different with $v_{Ga}^{3-}$ attracting $As_{i}^{1+}$ more than the other vacancies. To account for the increasing Coulombic attraction between $As_i$ and the vacancy types, effective reaction radii were set to $1.0\times10^{-10}$, $1.0\times10^{-8}$, and $1.0\times10^{-6}$~ cm for Reactions 1, 2 and 4, respectively.

\begin{figure}[h!]
    \centering
    \includegraphics[width=\columnwidth]{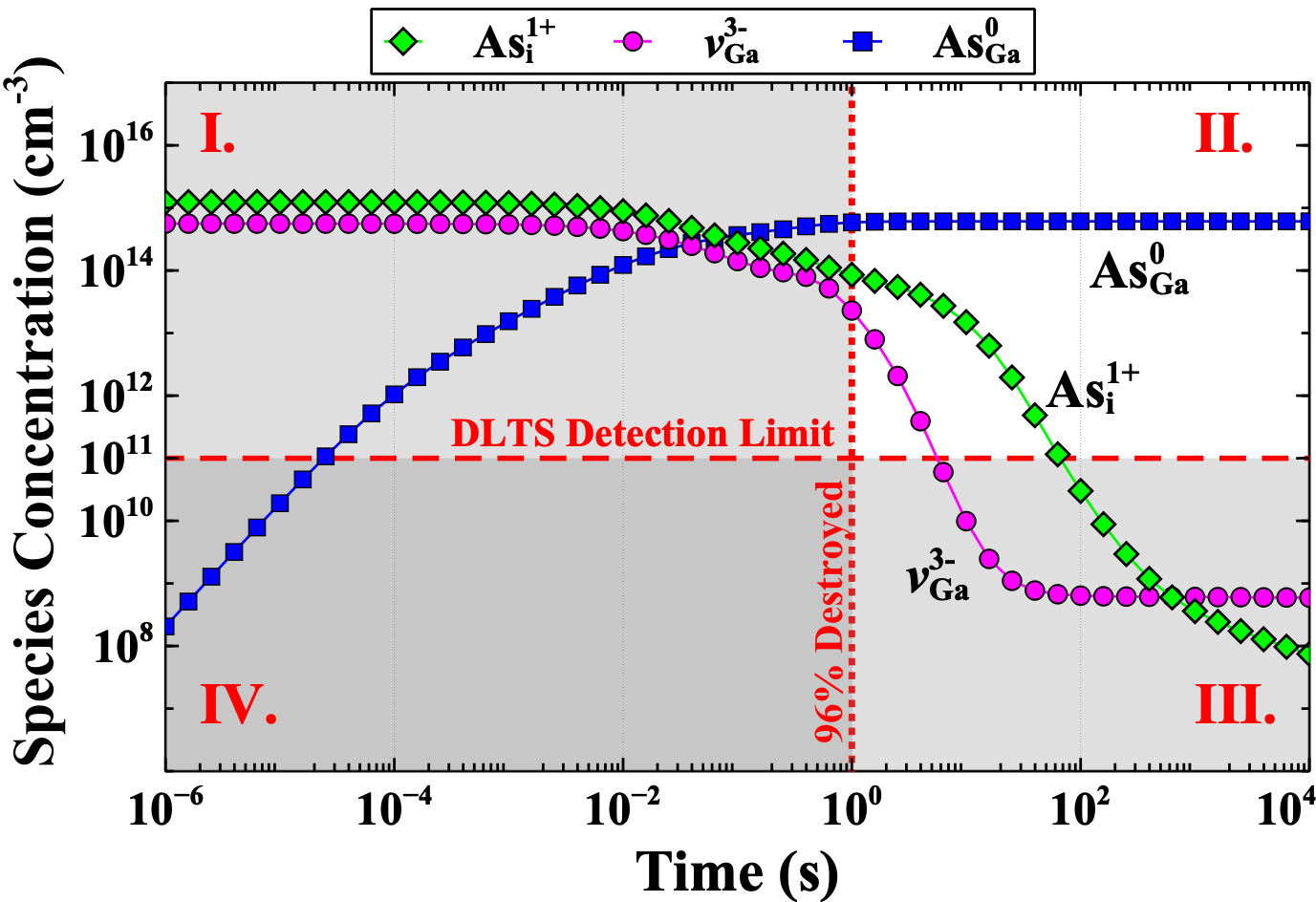}
    \caption{Species concentration for $As_{i}^{1+} + v_{Ga}^{3-}$ reaction. Red lines indicate DLTS concentration limit (dashed: $10^{11}$ cm$^{-3}$) and when 96\% of $v_{Ga}^{3-}$ is annihilated (dotted: 1~s). Intermediate steps are rapidly occurring and only $As_{Ga}^0$ is shown.}
    \label{fig:fig4}
\end{figure}

In Fig. \ref{fig:fig4}, the $As_i^{1+}+v_{Ga}^{3-}$ reaction is shown. After charge equilibration, the strong Coulomb attraction (Region I) results in $v_{Ga}^{3-}$ population collapsing below $10^{11}$~cm$^{-3}$ after 5~s and below $10^9$ cm$^{-3}$ after 60~s (Region III) where its population flattens. Flattening occurs because as $v_{Ga}^{3-}$ population declines, its probability of reacting with $As_i^{1+}$ gets overtaken by the largely more populated $v_{Ga}v_{As}^{2-}$. The $As_i^{1+}$ also experiences a diminishing population because, unlike $v_{Ga}^{3-}$, it is consumed by $v_{Ga}$ reactions and charge re-equilibration causing its charge state to change to $1-$. Formation of $As_{Ga}^0$, denoted by an increasing population (increasing slope), reaches its peak after about 5~s where its slope flattens in Region II. This rapid $v_{Ga}^{3-}$ decline provides a direct explanation for its invisibility to experimental probing.

Despite DLTS revealing several defects in GaAs, none have been identified as the Ga vacancy. Generally, DLTS includes filling active defects with a voltage pulse and measuring gradual changes in capacitance---recording capacitance transients. Conventional DLTS setups have concentration detection limits described by $N_{defect}/N_{doping}\approx \delta C_{max}/C_0$ where $N$ and $C$ denote concentration and capacitance, respectively \cite{schroder2005}. For our system (Si-doping $=1.00\times10^{15}$ cm$^{-3}$), conventional DLTS sensitivity ($\delta C/C_0\approx 10^{-5}-10^{-4}$ \cite{claeys2002}) is expected to be $N_{defect}=(\delta C_{max}/C_0)N_{doping}\approx10^{11}$~cm$^3$ (red dashed line in Fig. \ref{fig:fig4}). This limit is roughly 100 times larger than our predicted $v_{Ga}$ density ($<10^9$ cm$^3$), making it impossible to detect through DLTS. Including a high-sensitivity bridge improves this sensitivity to $\approx10^{9}$ cm$^3$ ($\delta C/C_0\approx10^{-6}$) \cite{Misrachi1980}---still not sensitive enough to see the Ga vacancy. Moreover, inclusion of omitted (slower) $Ga_i^{1+}$-vacancy reactions, with Coulomb interactions similar to $As_i^{1+}$, would reduce $v_{Ga}^{3-}$ population further. Defect populations existing in Regions III or IV will not be seen by experimental probes.

The $v_{Ga}$ is increasingly difficult for DLTS to observe because its population is non-constant and rapidly changing as it plummets below detectable limits---it's effectively invisible. Therefore, any chemical behavior, such as $v_{Ga}$ population decline, which saw roughly 96\% annihilated after 1 s (red dotted line in Fig. \ref{fig:fig4}), will not be observed under current detection capabilities. The $v_{Ga}$ exists, predominately, in the immeasurable regimes (Regions I and III) of the DLTS technique. 

\begin{figure}[t]
    \centering
    \includegraphics[width=\columnwidth]{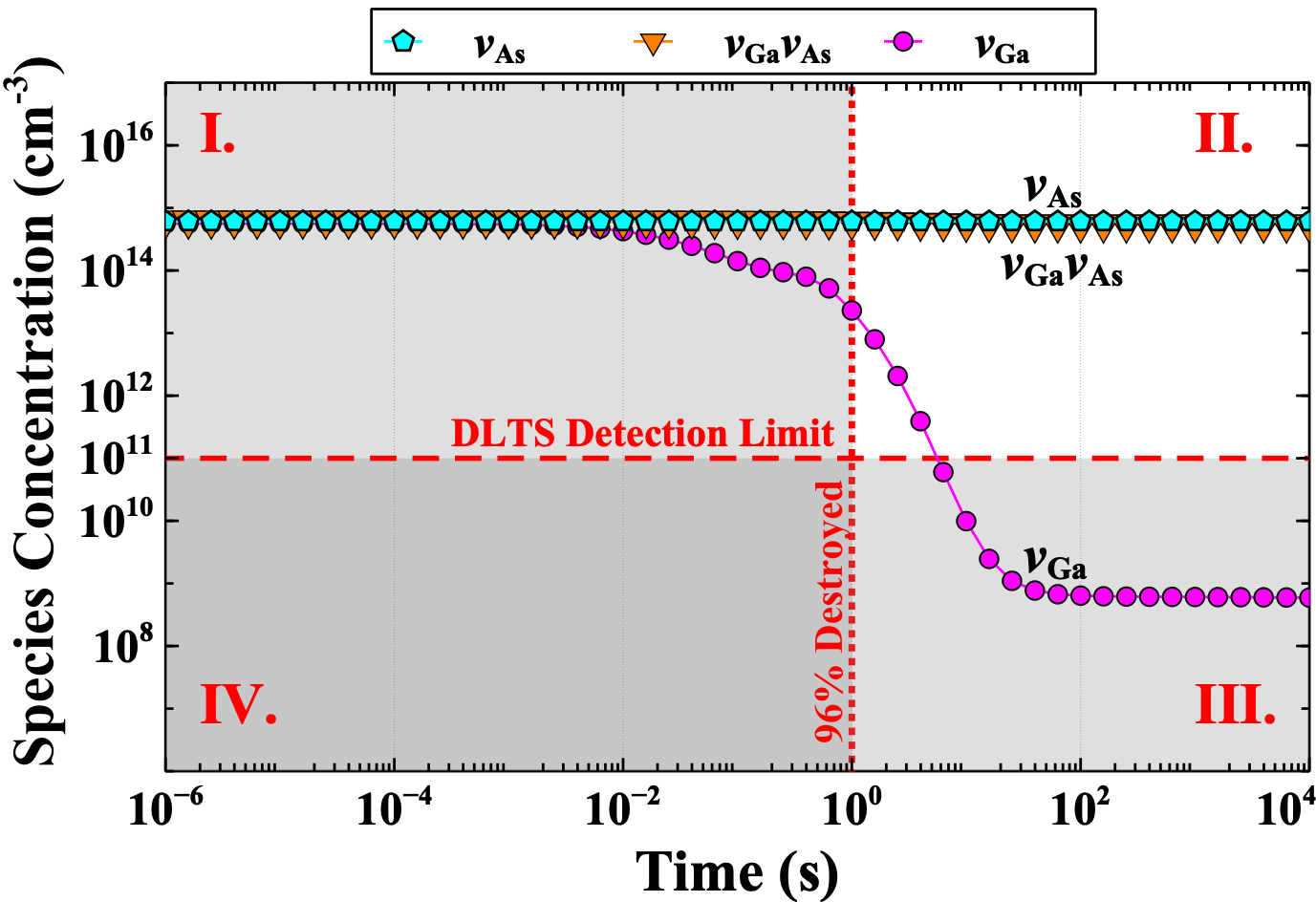}
    \caption{Total species concentrations for $v_{As}$, $v_{Ga}v_{As}$, and $v_{Ga}$. Red lines indicate DLTS concentration limit (dashed: $10^{11}$ cm$^{-3}$) and when 96\% of $v_{Ga}$ is annihilated (dotted: 1 s).}
    \label{fig:fig5}
\end{figure}

Total populations (sum of all defect charge states) for each vacancy type ($v_{Ga}$, $v_{Ga}v_{As}$ and $v_{As}$) are plotted over the duration of $As_i^{1+}$-vacancy reactions in Fig.~\ref{fig:fig5}. The total $v_{Ga}$ population experiences a collapse resembling that of its $3-$ charge state. Most of $v_{Ga}$ exists as $v_{Ga}^{3-}$, the source of the mid-gap Fermi level pinning. In contrast to $v_{Ga}$, $v_{Ga}v_{As}$ and $v_{As}$ populations do not collapse; instead, their populations remain largely unaffected and survive. Recall, $v_{Ga}v_{As}$ and $v_{As}$ have been observed as E1-E2 \cite{schultz2015} and E3a \cite{schultz2022}, respectively. Also, E1-E2 and E3 do not anneal until temperatures above 490 K \cite{pons_bourgoin1985, taghizadeh2018} suggesting that at room-temperature $v_{Ga}v_{As}$ and $v_{As}$ should maintain substantial populations and survive any early-time defect chemistry. In agreement with observation, our simulation predicts that $v_{Ga}v_{As}$ and $v_{As}$ survive the Coulomb-driven wave of $As_i$ reactions. Their survival provides further reassurance of the simulations physical accuracy and confirms $v_{Ga}$ preferential annihilation.

Simulations in Figs. \ref{fig:fig4} and \ref{fig:fig5} were allowed to run for $10^4$ s. During this time, $As_i$ becomes $As_i^{1-}$ and begins repelling all negatively charged vacancies. Currently, we are unable to capture this dynamical change that must occur during this later stage. As a result, our model represents physical reality up to roughly $10^2$ s when $As_i^{1+}$ population drops below detectable limits. The extended timescale demonstrates the AIDE method's potential and aptitude for exploring experimentally inaccessible regimes. 

In summary, the physical explanation to the long-standing mystery of the missing $v_{Ga}$ in irradiated GaAs is simple and is resolved once dynamical effects are deliberately considered in a multiscale atoms-to-devices analysis. This innovative approach revealed that $v_{Ga}$ is not observed because it has been preferentially annihilated---below detectable levels---by highly mobile $As_i$ in a rapid initial wave of defect reactions. This confirms the hypothesized stability of a $v_{Ga}^{3-}$ deep in the band gap and validates the DFT-predicted ionization of $As_i$ into a positive ion at mid-gap. Interestingly, the lack of experimental evidence for $v_{Ga}$ is, in fact, evidence for its absence. It is not due to some undiagnosed experimental blindness ($v_{Ga}$ is not there to be seen), nor to errors in DFT predictions (deducing the fate of $v_{Ga}$ depends on DFT defect level calculations being accurate), nor the result of some unknown, unusual kinetic processes---the initial annealing kinetics can be modeled effectively in a defect device model using well-established defect physics.

In addition to providing an explanation for the experimentally invisible $v_{Ga}$, the AIDE method provides insight into the rich defect physics of regimes that are inaccessible to experimental probing. The AIDE method will serve as a virtual experiment to bound estimates for difficult-to-measure quantities such as capture cross-sections and diffusion activation energies. 

\noindent \textbf{Acknowledgements} We thank G. Vizkelethy, J. M. Cain, B. A. Aguirre, E. S. Bielejec, and P. J. Griffin for useful discussion. We are indebted to G. Vizkelethy and W. R. Wampler for critical comments on the manuscript. Sandia National Laboratories is a multimission laboratory managed and operated by National Technology \& Engineering Solutions of Sandia, LLC, a wholly owned subsidiary of Honeywell International Inc., for the U.S. Department of Energy’s National Nuclear Security Administration under contract No. DE-NA0003525. This work was supported by a Laboratory Directed Research and Development (LDRD) project (No. 229430, SAND2025-04630O).

\bibliography{aps}

\section{End Matter}
\noindent The REOS suite is a device theory simulation package capable of implementing and running damage models by iteratively solving the classical continuum semiconductor transport equations \cite{shockley1949,roosbroeck1950} for a collection of species in an idealized device structure. The suite obtains the electrical response of the one-dimensional sample connected to an electrical circuit. In this work, the one-dimensional slab of length $2.1\times10^{-4}$ cm (simulation space) is composed of imperfections (defect and dopant species) and charge carriers ($e^-$/$h^+$).

\textit{Defect Population Size Criteria.} Experimentally reported defect concentrations for radiation-induced GaAs defects are sparse and never comprehensive. Defect populations were chosen by considering experimental and theoretical observation and intuitive reasoning. Criteria:
\begin{enumerate}
    \item Single vacancy populations should be equal ($v_{Ga}=v_{As}$).
    \item The divacancy population must be larger than single vacancies $v_{Ga}v_{As}>v_{Ga}=v_{As}$ because the experimentally reported E1-E2 are larger than E3 \cite{pons1980}.
    \item Based on the manner that these defects get created---an interstitial leaves behind a vacancy \cite{frenkelpair}---the total number of vacancies should be roughly equal to the number of interstitials ($v_{Ga}+v_{As}+2v_{Ga}v_{As}= As_i+Ga_i$ or $v_{As}+v_{Ga}v_{As}= As_i$ and $v_{Ga}+v_{Ga}v_{As}= Ga_i$).

    \item Fermi level pinning implies that there are equal (or more) available states for dopant electrons ($e$), i.e., the number of available negatively charged vacancy states must be equal (or greater) than the number of Si-dopant electrons $e^{v_{Ga}}+e^{v_{As}}+e^{v_{Ga}v_{As}} > e^{Si_{Ga}}$.
    \item The Ga vacancy consumes the majority of the available electrons ($e^{v_{Ga}} > e^{v_{As}}, e^{v_{Ga}v_{As}}$).
\end{enumerate}  
Using this criteria, defect population sizes were chosen to be $As_i=Ga_i=1.24 \times 10^{15}$, $v_{As} = v_{Ga} = 6.10 \times 10^{14}$, and $v_{Ga}v_{As} = 6.30 \times 10^{14}$ cm$^{-3}$. 

Increasing (decreasing) the number of interstitials increases (decreases) the probability that $As_i^{1+}$ will react with vacancies, accelerating (decelerating) the reaction. Slightly increasing (decreasing) the number of vacancies shifts the Fermi level deeper (shallower) in the band gap, resulting in more (less) $As_i^{1+}$ interstitials accelerating (decelerating) the reaction. Too many vacancies shift the Fermi level deeper changing the dominant defect charge states which results in new favorable defect reactivity.  

\textit{Reactions Description.} The reactive-transport equation (RTE) governs the temporal evolution of chemical species $c_i(\textbf{r}, t)$ as they participate in chemical reactions ($e^-/h^+$ capture/emission reactions and defect-defect reactions). The RTE is given by

\begin{equation}
\label{eqn:rte}
    \frac{\partial c_i(\textbf{r}, t)}{\partial t}  = \sum_j \nu_{ij} r_j + \nabla \cdot (c_i \frac{D_i}{k_BT})\nabla \Phi_i
\end{equation}
where the first term on the right-hand side describes the reaction rate $r_j$ for reaction $j$ at a stoichiometric coefficient $\nu_{ij}$. The second term describes transport (drift-diffusion) and is composed of the diffusion coefficient $D_i=D_0e^{-E_m/k_BT}$ where $D_{0i}$, $E_{mi}$, $k_B$, and T are the diffusion prefactor, migration barrier, Boltzmann constant, and temperature, respectively. The electrochemical potential is defined by $\Phi_i=z_i\phi+\mu_i$ where $z_i$ is the species charge, $\phi$ is the electric potential calculated from the Poisson equation, and $\mu_i$ is the chemical potential. The $D_i$ and $r_j$ are highly important and will be for charge carrier reactions and defect-defect reactions. 

\textit{Charge Equilibration (CE).} To investigate the Fermi level shift, an initial defect population was given to each neutral defect and allowed to distribute throughout the system, via $e^-$/$h^+$ capture and emission (charge carrier) reactions, until achieving equilibrium. Charge carrier reactions enable charge flow while also maintaining charge neutrality via the charge-conserving reactions $a^{1-}+h^+\rightleftarrows a^0$ and $a^0+e^-\rightleftarrows a^-$ where $a$ is a generic defect. Charge equilibration depends only on charge carrier reactions---no defect-defect reactions---and occur rapidly because $e^-$ and $h^+$ (diffusion coefficients of 207 and 11~cm$^2$/s, respectively \cite{neuberger}) diffuse much faster than defects. The reaction rate for the generic $h^+$ capture reaction discussed above is given by
\begin{equation}
    r_j^{a^{1-}}= v_{th}\sigma_j\Big\{-[a^{1-}]n_he^{-E_{f}/kT}+N_V[a^0]e^{-E_{r}/kT}\Big\}
\end{equation}
where $v_{th}$, $\sigma$, $n_h$, are the thermal velocity, capture cross-section, and number of holes, respectively. The $E_{f}$ and $E_{r}$ are activation energies for the forward and reverse reaction---DFT-predicted defect levels. With minor changes, a similar equation can be written for $e^-$ capture. As seen in Fig. \ref{fig:fig1}, the defects exist in positive, neutral, and negative charge states creating an imminent Coulomb interaction between charge carriers and defect charge states. Since it is generally accepted that Coulombic interactions change the size of capture cross-section \cite{bonch1968}, cross-sections were set to range from $10^{-11}$ to $10^{-18}$~cm$^2$ for attractive and repulsive Coulomb interactions, respectively. For capture by a neutral defect species, cross-sections were set to $10^{-15}$~cm$^2$. Chosen values for each reaction are given in the Supplemental Tables 1-7. There is substantial uncertainty in the capture cross-sections with widely ranging reported values. However, these values agree with a range of experiments for charged (attractive or repulsive) and neutral interactions \cite{Benton_1982,Kono,taghizadeh2018}. This representation is both of correct order and reflects the varying Coulomb interactions.

\textit{Defect-Defect Reactions: Diffusion-Controlled.} The $As_i$ dominated all defect-defect reactions. The determination of its parameters had immense importance to this work. The diffusion prefactor is known to vary widely among materials and material defects \cite{pichler}. Here, $D_0=10^{-2}$ cm$^2$/s, similar to other device simulations \cite{wampler2008,wampler2015}, was nominally chosen. The $As_i^+$ diffuses athermally via the Bourgoin-Corbett mechanism, i.e., its motion through the lattice occurs by changing its charge state by alternating capture of electrons and holes \cite{bourgoin-corbett}. Its thermal migration barrier was set to 0.40 eV (20\% lower than 0.5 eV) to mimic this athermal behavior. Using only the thermal barrier 0.5 eV---instead of 0.40 eV---produced similar results with $v_{Ga}$ density plummeting to immeasurable scales after $\approx$200 s instead of $\approx$5 s. Interestingly, even with the thermal migration barrier, $v_{Ga}$ would still be difficult to experimentally observe. Diffusion coefficients for all immobile species were set to zero. The $Ga_i$ diffusion coefficient was also set to zero since the reactivity in our simulation is dominated by $As_i$. 

Defect-defect reactions take the form $\alpha A+\beta B\rightleftarrows \gamma C+\delta D$, where A, B, C, and D are generic defects and $\alpha$, $\beta$, $\gamma$, and $\delta$ are their coefficients. The forward reaction rate for $j$ diffusion-controlled reactions ($r_{jf}^d$) takes the form
\begin{equation}
    r_{jf}^d=k_{jf}^d[A_j]^{\alpha_j}[B_j]^{\beta_j}
\end{equation}

\begin{equation}
    k_{jf}^d = 4\pi R_{eff} [D_{A}+D_{B}]
\end{equation} 
where $R_{eff}$ is an effective reaction radius of reaction $j$ and $D_A$ and $D_B$ are diffusion coefficients of reactant defects A and B. Forward reaction radii were used to simulate the Coulomb attraction between $As_i^{1+}$ and the negatively charged vacancies. Reaction radii are difficult to determine and were set on the order of a lattice constant $10^{-8}$ cm: $1.0\times10^{-10}$, $1.0\times10^{-8}$, and $1.0\times10^{-6}$~cm for Reactions 1, 2 and 4, respectively. Reverse reactions were not included in this work and are not discussed.

\textit{Defect-Defect Reactions: Transition-State Controlled.} For each reaction, a complex forms before $As_i$ hops into a vacancy site, i.e., $As_i$ must overcome an energy barrier ($E_b$) before occupying the vacancy site. This transition-state must be included as they occur after Diffusion-controlled (Coulomb-driven) reactions. The forward transition-state reaction rate ($r_{jf}^{ts}$) is given by

\begin{equation}
    r_{jf}^{ts} = \frac{kT}{h}e^{-E_b/kT}
\end{equation}
where $h$ is the Planck constant. For the energy barrier ($E_b$), physically reasonable values of 0.75, 0.50, and 1.00~eV were chosen for Reactions 1, 2, and 4. Note, when all energy barriers are set to 1 eV, nearly identical results persist. Further details are provided in the Supplemental Material.

Several parameters used in device simulations are not well-known, e.g., capture cross-sections, diffusion rates, and reaction radii, both experimentally and theoretically. The chosen parameters were physically reasonable and produced results that agree with well established defect physics.

Further details on the methods used in the REOS suite will be provided in a separate work. 

\newpage

\begin{figure} [h!]
    \centering
\includegraphics[width=\textwidth]{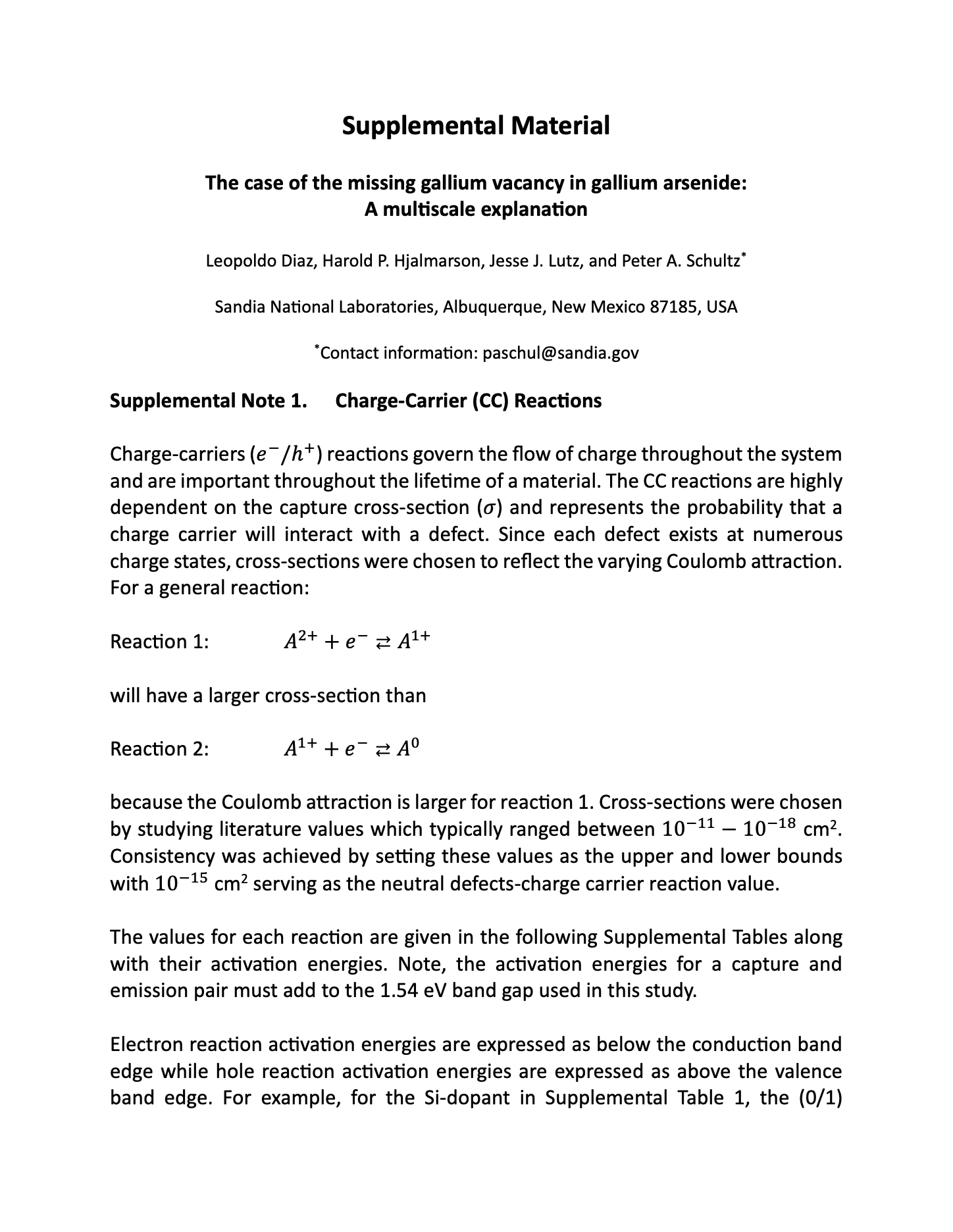}
\end{figure}

\begin{figure} [h!]
    \centering
\includegraphics[width=\textwidth]{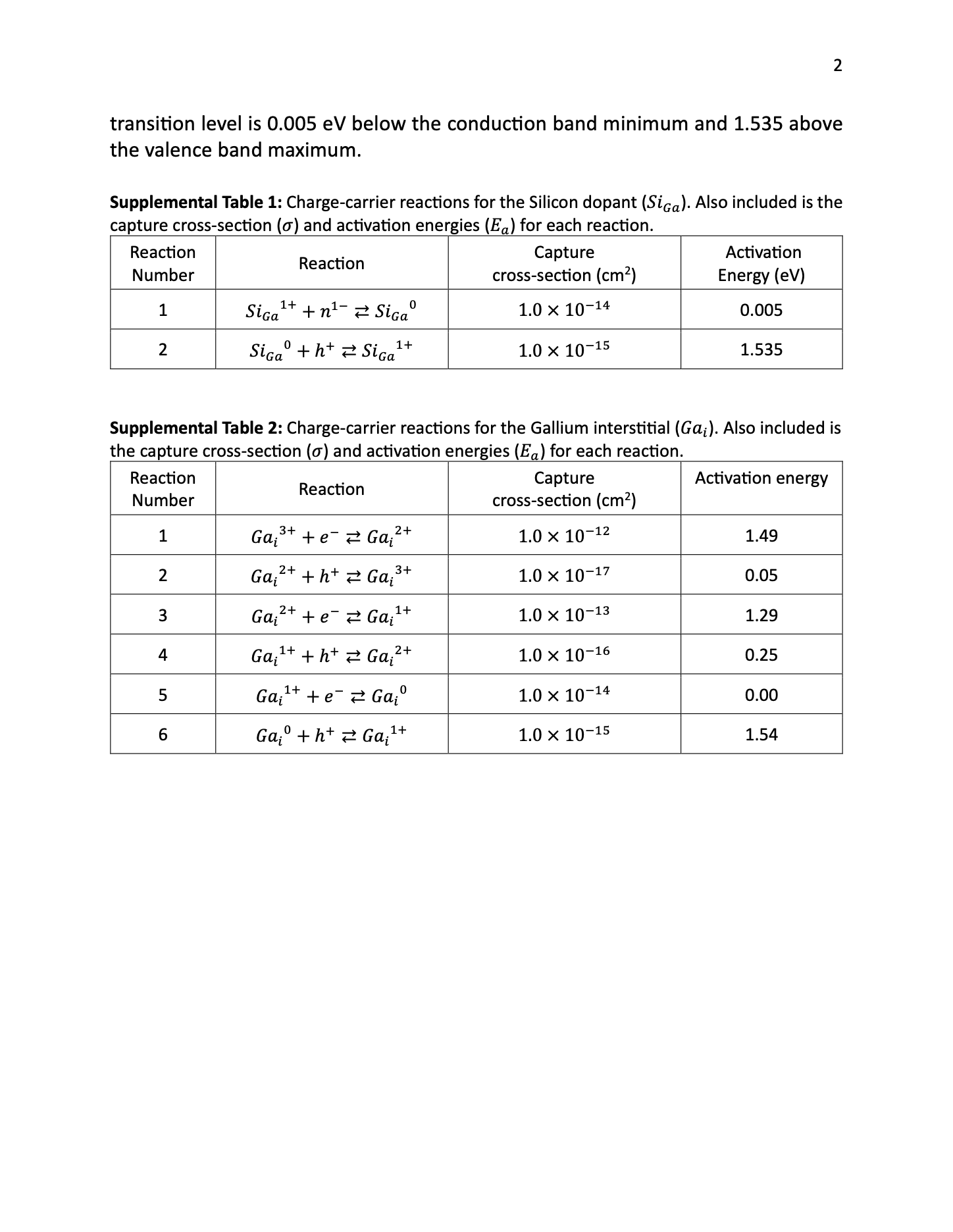}
\end{figure}

\begin{figure} [h!]
    \centering
\includegraphics[width=\textwidth]{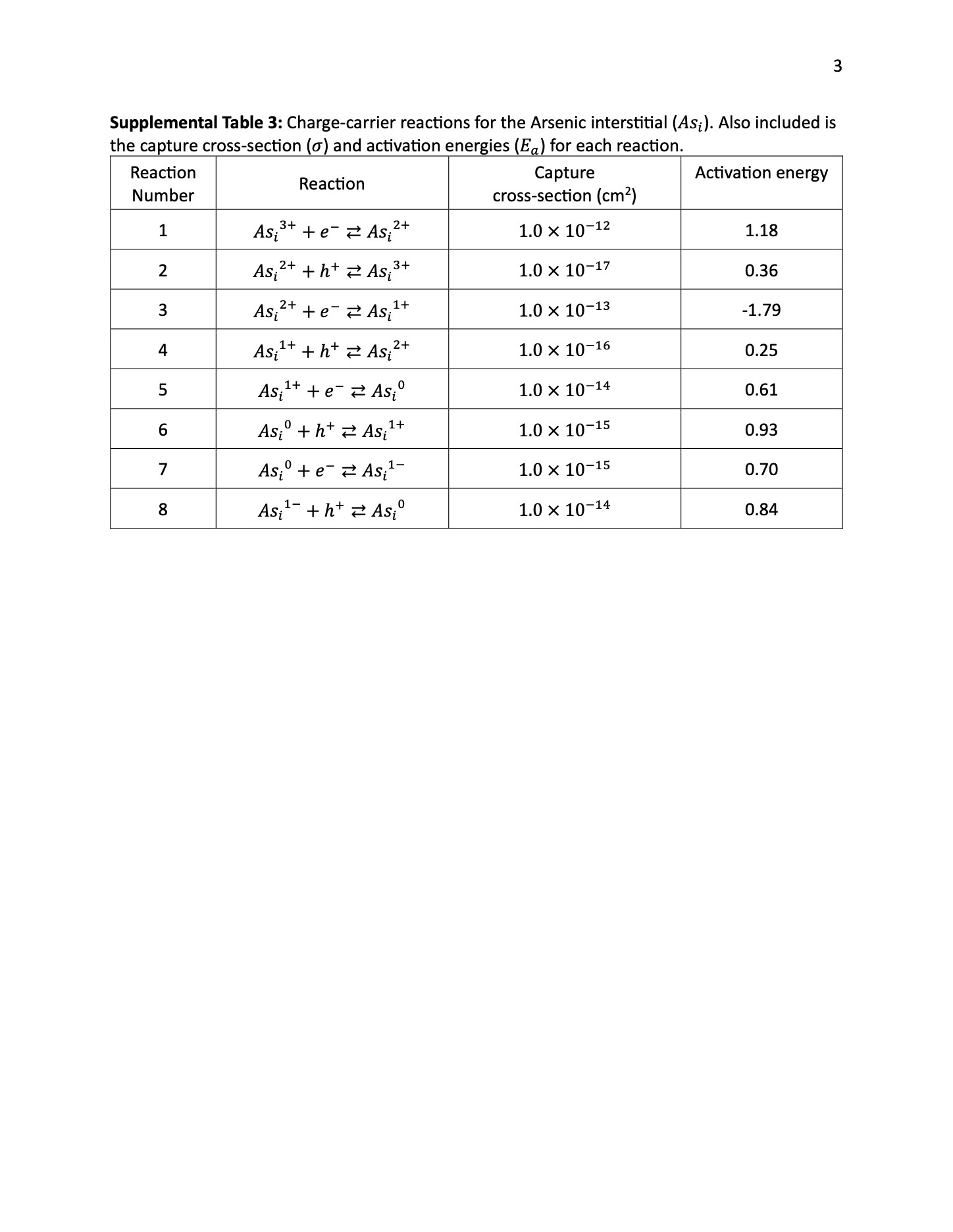}
\end{figure}

\begin{figure} [h!]
    \centering
\includegraphics[width=\textwidth]{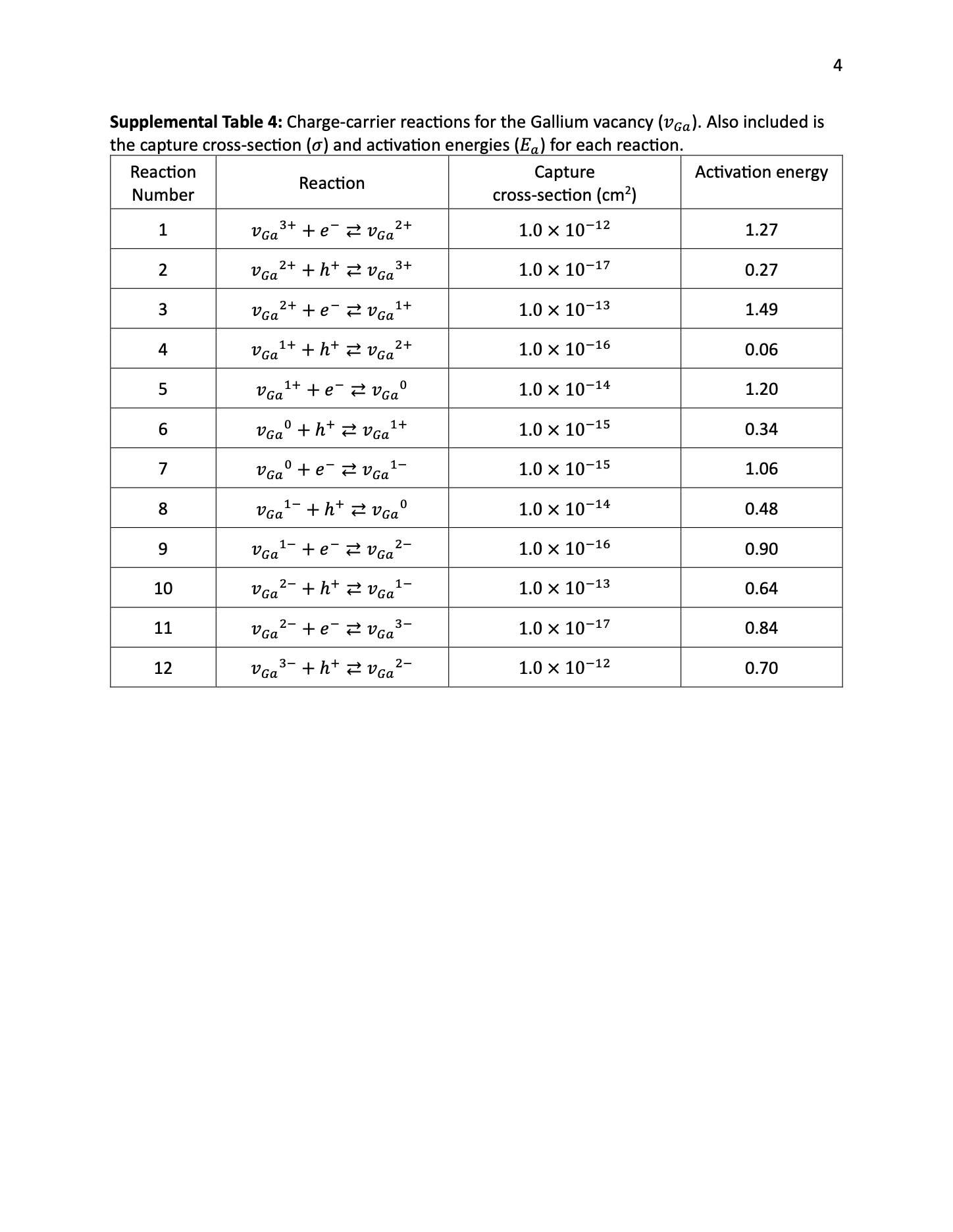}
\end{figure}

\begin{figure} [h!]
    \centering
\includegraphics[width=\textwidth]{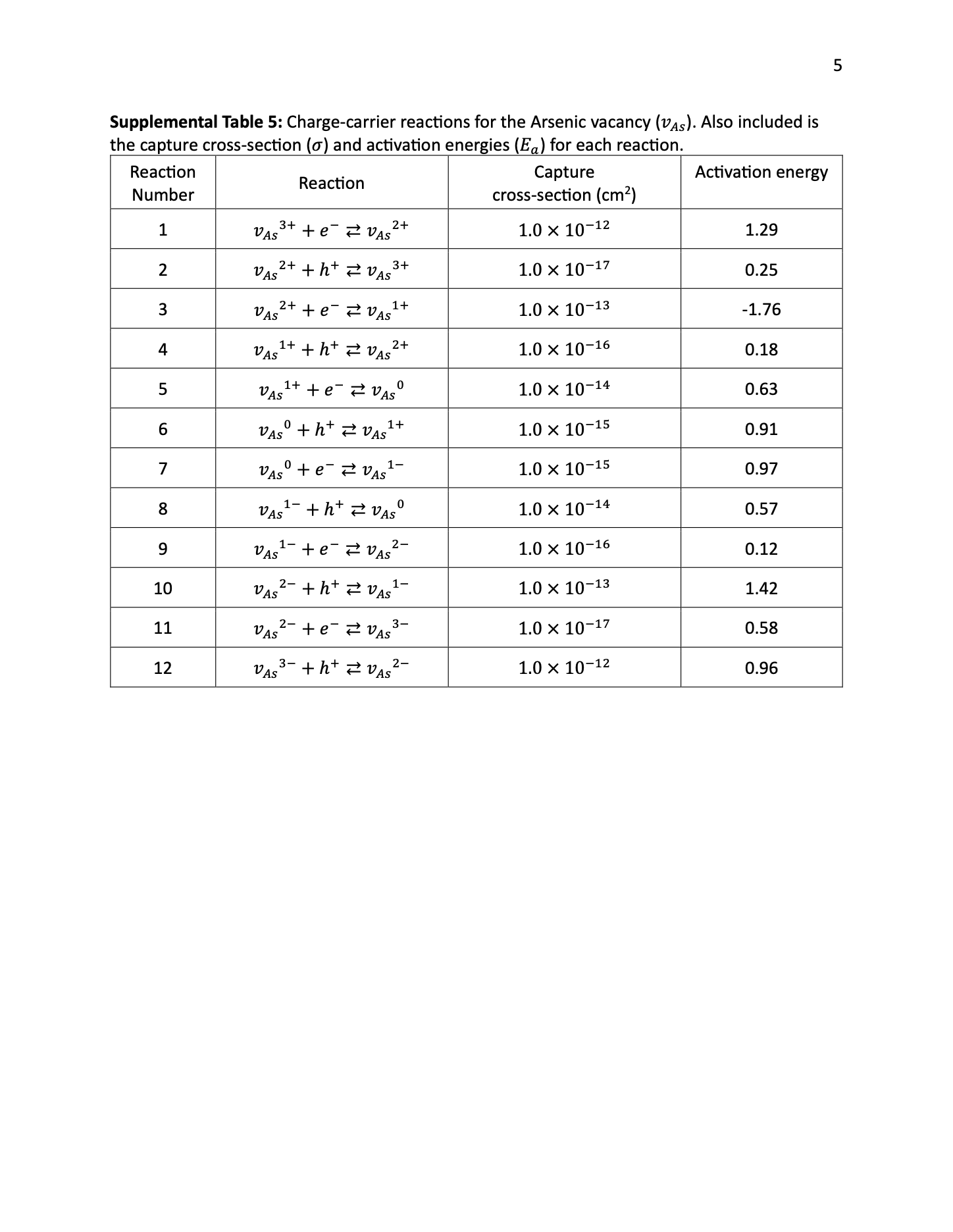}
\end{figure}

\begin{figure} [h!]
    \centering
\includegraphics[width=\textwidth]{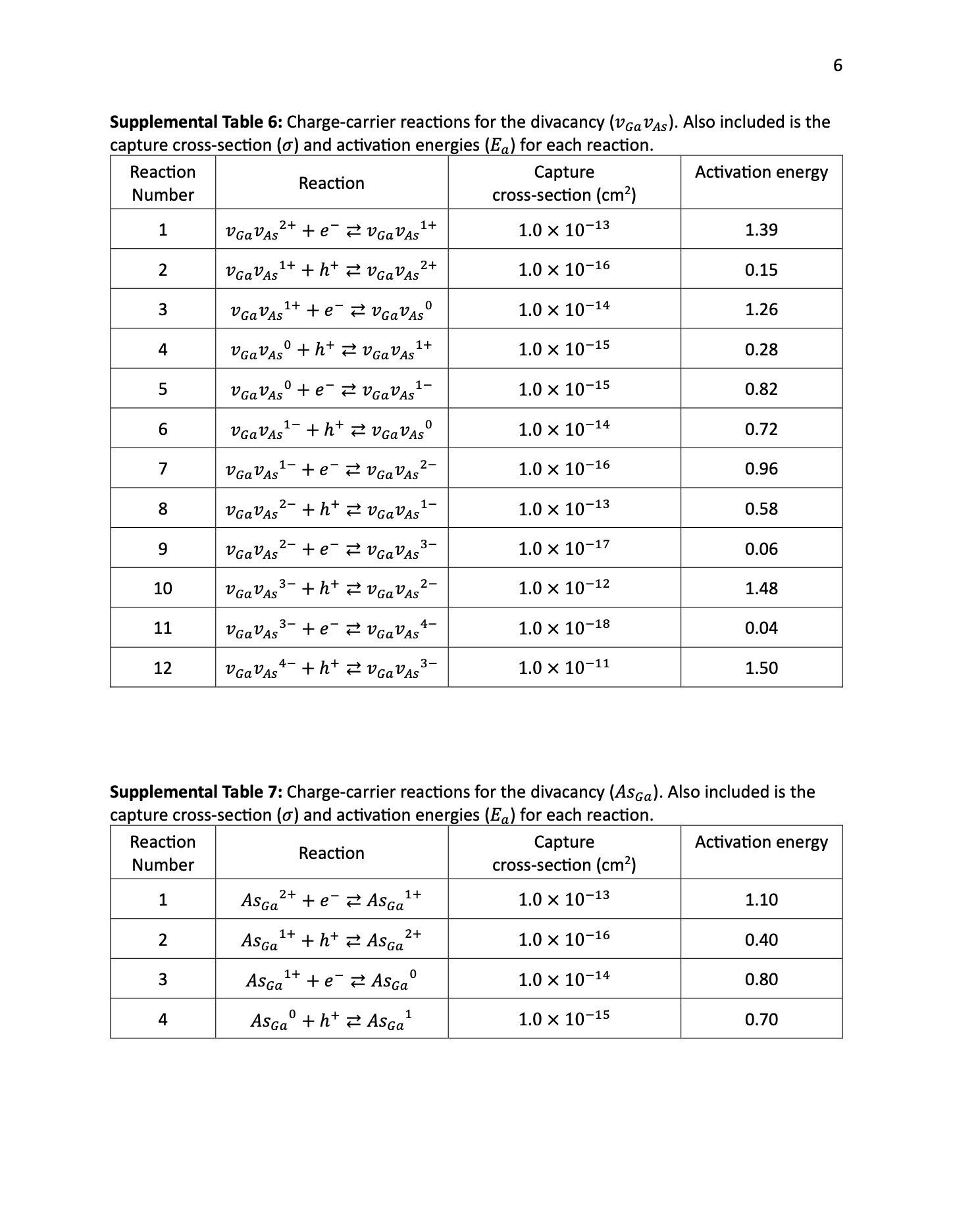}
\end{figure}

\begin{figure} [h!]
    \centering
\includegraphics[width=\textwidth]{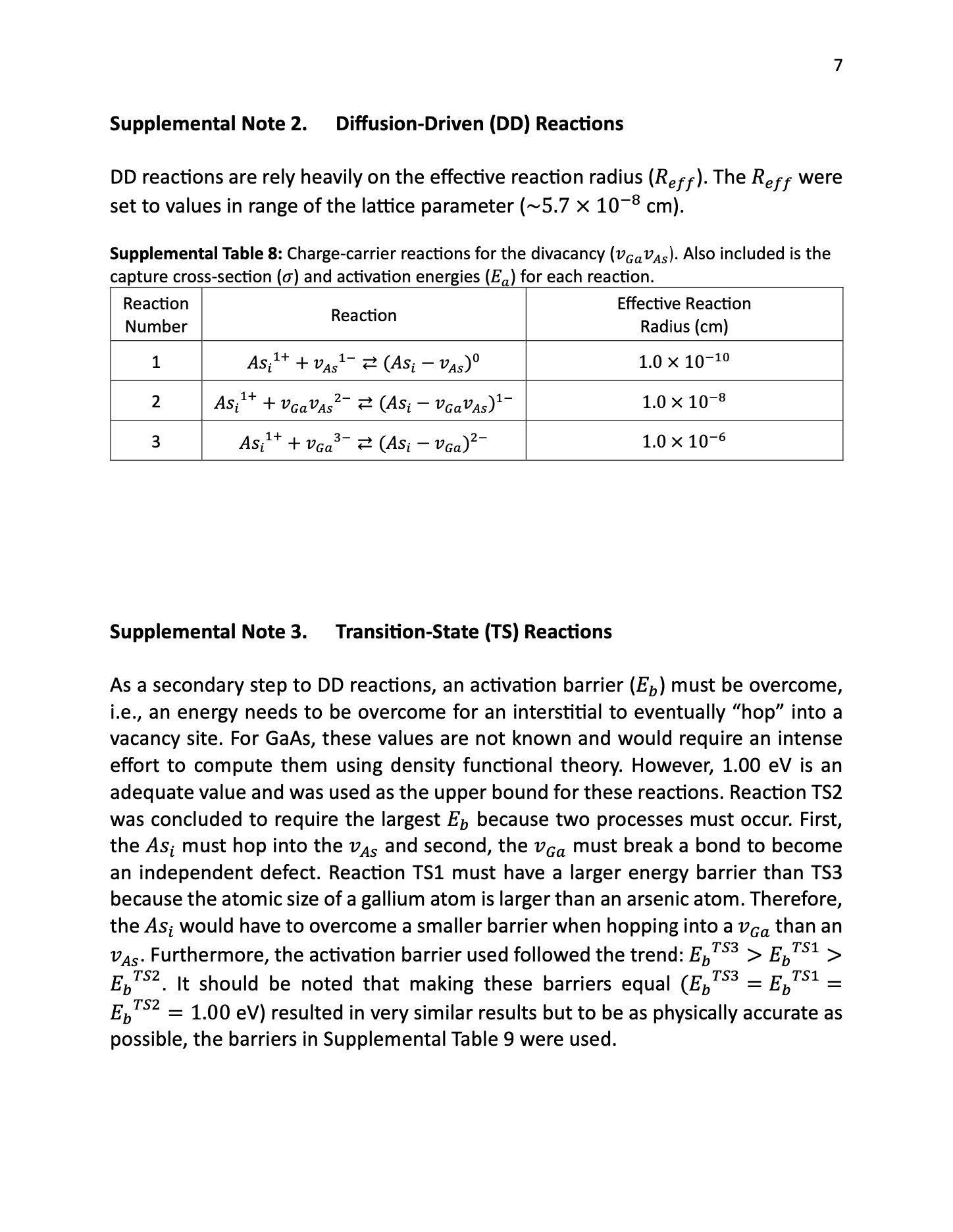}
\end{figure}

\begin{figure} [h!]
    \centering
\includegraphics[width=\textwidth]{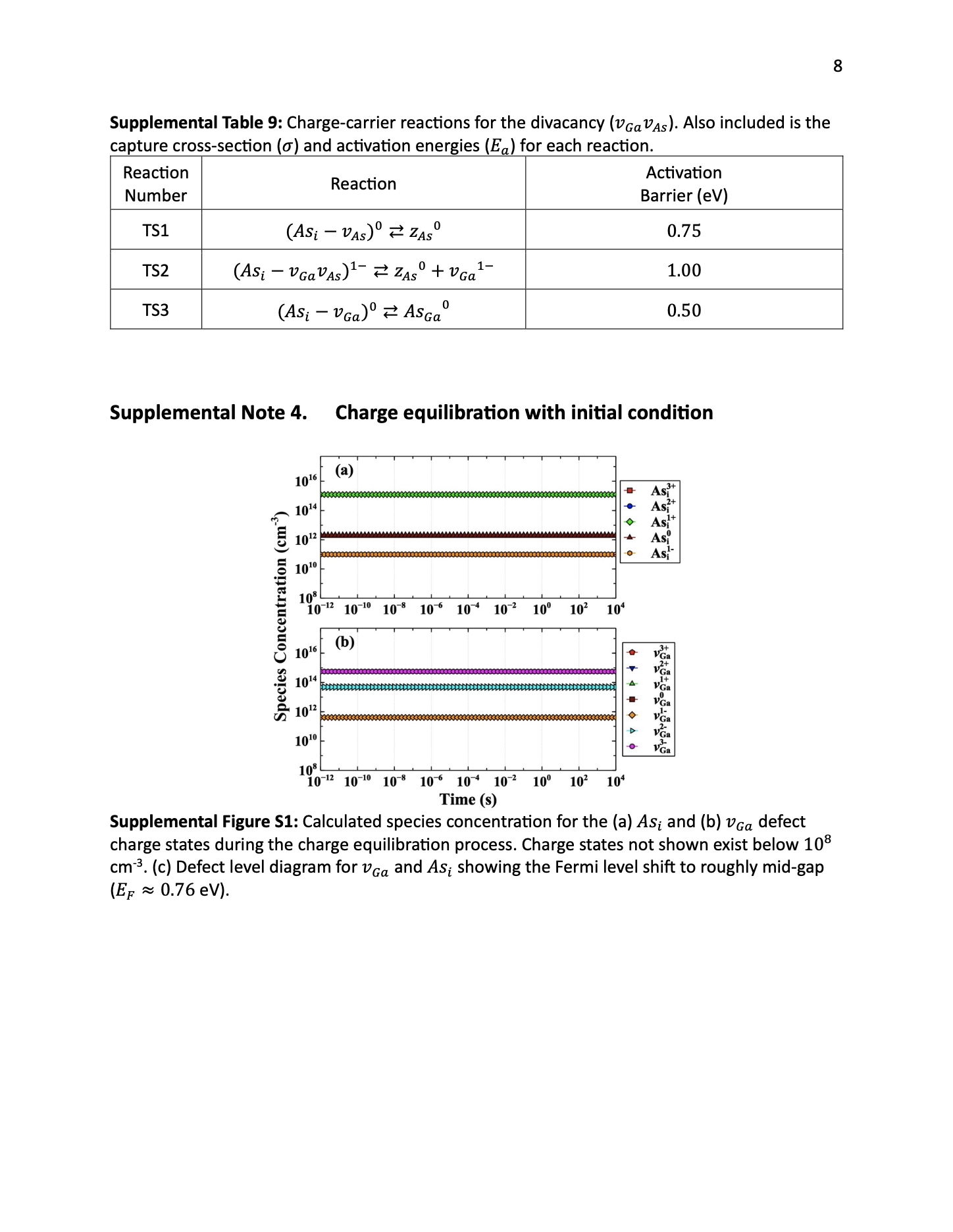}
\end{figure}

\end{document}